\documentclass[amsmath,amssymb,aps,twocolumn,pre]{revtex4-2}
\usepackage{graphicx}
\usepackage{dcolumn}
\usepackage{bm}
\usepackage{array}
\usepackage{booktabs}
\usepackage{physics}
\usepackage{verbatim}

\begin{document}

\preprint{APS/123-QED}

\title{Quantum Ising Spin-glass Otto engine}
\email[Correspondence email address: ]{asozdemir@ku.edu.tr}
\author{Asl{\i} Tuncer$^\text{1*}$ and Batu Yal\c{c}{\i}n$^\text{2}$}
       \affiliation{$^\text{1}$Ko\c{c} University, Institute of Physics, Sar{\i}yer, 34450, \. Istanbul, T\"urkiye\\ $^\text{2}$American Robert College of Istanbul, Be\c sikta\c s, 34345, \. Istanbul, T\"urkiye}      
    
\date{\today}

\begin{abstract}
We investigate a quantum Otto engine with a quantum Ising spin glass as the working medium to explore the scaling behavior of work output and thermodynamic performance concerning system size, particularly near the critical point. Specifically, we explore the two operating modes of the Otto engine, namely the heat engine and refrigerator modes. We observe a double-peaked structure in the heat engine regime, leading to superlinear scaling in both work output and thermodynamic performance near the critical point. Additionally, in the refrigerator regime, superlinear scaling in refrigerator efficiency can be achieved at high and low temperatures, significantly outperforming models with uniform Ising interactions. These findings suggest that disorder and frustration in quantum Ising spin-glass systems could significantly impact thermodynamic performance in quantum heat engines and refrigerators, potentially opening up new avenues for improvement.
\end{abstract}


\maketitle


\section{\label{sec:Intro}Introduction}

Investigation of quantum analogs of the Otto engines has garnered significant attention. This interest is motivated by the pronounced impacts of quantum phenomena such as quantum fluctuation \cite{Campisi_2015}, superposition \cite{Ou2013, supposCarnot}, entanglement \cite{QheCarnotPow}, and coherence \cite{Scully2002} on these cycles.
The literature on these thermodynamic cycles is mostly focused on quantum heat engines and quantum refrigerators~\cite{kosloff2014,kosloff2017,tuncer,Kurizki2022,deffner}. Numerous theoretical investigations of these systems are available \cite{Kumar2023, RAlicki_1979}, often dealing with simple few-level and/or single-particle quantum thermodynamic cycles \cite{Varinder2020, kieu2005quantum, CarlMBender_2000, Narevicius_2009, kosloff2014, bhattach}. Recent studies have extended to more intricate systems, encompassing interacting particles \cite{Bush2023}, atomic collisions \cite{collision}, strong coupling with baths \cite{Kaneyasu_2023}, and inner friction \cite{Alecce_2015}. Our work aligns with these trends, where computational treatments of multi-level spin systems with larger system sizes have been explored \cite{Piccitto_2022, Alecce_2015}. The impact of spin frustration on the efficiency of thermodynamic cycles has also been investigated \cite{Azimi_2014}. Nevertheless, the exploration of quantum spin glasses, specifically magnetic states manifesting exotic behavior due to system frustration and demonstrating pronounced quantum fluctuations \cite{Rieger2005}, with substantial system sizes remains an unexplored domain. 

In this study, we investigate the potential use of the quantum Ising spin-glass $1d$-chain as a working substance in a quantum Otto engine to enhance performance. This is the simplest nontrivial interacting
quantum model with quenched randomness. We also show that superlinear scaling of the performance emerges when the working substance is on the verge of a phase transition.

As the working substance approaches the critical point in a quantum Otto engine, the enhancement in performance is a well-known phenomenon~\cite{Campisi2016vu, Piccitto_2022, Fogarty_2021}. In this context, the impact of various exotic behaviors, such as the slowing down of the system's dynamics, the breaking of time-reversal symmetry, the formation of islands in the system, and the increase in long-range correlations between neighboring regions, on the engine's performance is significant~\cite{Benenti2011}. 

Moreover, the slow relaxation appears in the quantum Ising spin-glass model around the critical point $h_c$, i.e. critical quenched field. This slow relaxation behavior is called the Griffith singularity~\cite{Fisher1992,Fisher1995}. From this point of view, we examined our system to investigate the phase transition points and the effects on engine performance.


The paper is organized as follows: In Sec.~\ref{subsec:qsg}, we briefly introduce the quantum Ising spin-glass model and Griffiths phases. In section~\ref{sec:QOtto}, the four-stroke quantum Otto engine is described followed by the mathematical model, and the working regimes are presented. In Sec.~\ref{sec:res}, we will present our computational results on the thermodynamic performance of the Ising spin-glass QOE, in the context of the phase transitions. Finally, in Sec.~\ref{sec:conc}, we draw our conclusions.

\section{\label{subsec:qsg}The Quantum Ising Spin-Glass Model as Working Medium}
We focus on a model
of N spin-$1/2$ quantum particles coupled via the quantum Ising chain in a time-dependent transverse magnetic field as the working substance of our engine. The Hamiltonian of this model is
\begin{equation}
    \hat{H}(t)=-\sum_{i=1}^N J_{i}\hat{\sigma}^x_i \hat{\sigma}^x_{i+1}
-h(t)\sum_i \hat{\sigma}^z_i,
\label{eq:TIMham}
\end{equation}
here $\hat{\sigma}_j^\alpha$ are Pauli operators ($\alpha = x, y, z)$ acting on
the $j$-th spin, $\hat{\sigma}_{N+1}^\alpha=\hat{\sigma}_1^\alpha$ enforces the periodic boundary conditions, and $h(t)$ is the strength of the external field at time $t$. The spins $\hat{\sigma}^x_i$ and $\hat{\sigma}^x_{i+1}$ interact with each other by coupling coefficients $J_{i}$ that obey a Gaussian distribution with zero mean $J_0=0$ and variance $1/N$. This model, the so-called Edwards-Anderson spin-glass model~\cite{SFEdwards_1975}, is similar to the Ising model, with the only interaction between nearest neighbor spins. Still, the difference is that they have randomly distributed ferromagnetic ($J_{i}>0$) and antiferromagnetic ($J_{i}<0$) interactions. We also set $\hbar=k_B=1$. 

In order to diagonalize the system Hamiltonian, we first apply the Jordan-Wigner transformation which maps equation~(\ref{eq:TIMham}) into the spinless-fermion Hamiltonian~\cite{Rieger2005}
\begin{align}
H(t)&=-\sum_{i=1}J_{i}(\hat{a}_i^\dag\hat{a}_{i+1} + \hat{a}_i^\dag\hat{a}_{i+1}^\dag +\text{h.c.})\nonumber \\
&+h(t)\sum_i (2\hat{a}_i^\dag\hat{a}_i-1),
\label{eq:1}
\end{align}
where $\hat{a}_i$ ($\hat{a}_i^\dag$) are fermionic annihilation (creation) operators. 
Note that we need to access the full eigenspectrum of the Hamiltonian in order to calculate the expectation values. However, this process is intractable for large system sizes $N$, as the size of the Hamiltonian grows exponentially. Thus, we use a special diagonalization trick called the Bogoliubov Transform~\cite{mbeng2020quantum}, which allows us to deal with a $2N \times 2N$ matrix instead of a $2^N \times 2^N$ one. More details can be found in Appendix \ref{sec:bog}.

While this model with random field interactions has not been studied in the context of quantum heat engines, it was extensively examined~\cite{Fisher1995,Fisher1998}, where several results on critical behavior and phase transitions were found, which will play a crucial role in our further analysis.

The critical point of the random transverse-field chain
is obtained by Shankar and Murthy~\cite{PhysRevB.36.536}. Then, following the method, the control parameters are
\begin{align}
    \Delta_h &= \overline{\ln{h}},\\
    \Delta_J &= \overline{\ln{J}}
\end{align}
where $J$ and $h$ are the variances of the probability distributions belongs to the $J_i$-couplings and $h_i$ random-fields, respectively~\cite{Fisher1995}. At the critical point,
\begin{equation}
    \Delta_h = \Delta_J = \overline{\ln{J}}\equiv \Delta_c,
    \label{eq:RTIMc}
\end{equation}
where the bar denotes the disorder average and $\Lambda_c$ is the critical value of the control parameters. 

Once the magnetic field is uniform, one can conclude that the quantum critical point is at  
\begin{equation}
    h_c = e^{\Delta_c} = e^{\overline{\ln{J}}}.
    \label{eq:hc}
\end{equation}

\begin{figure}
    \centering
    \includegraphics[scale = 0.5]{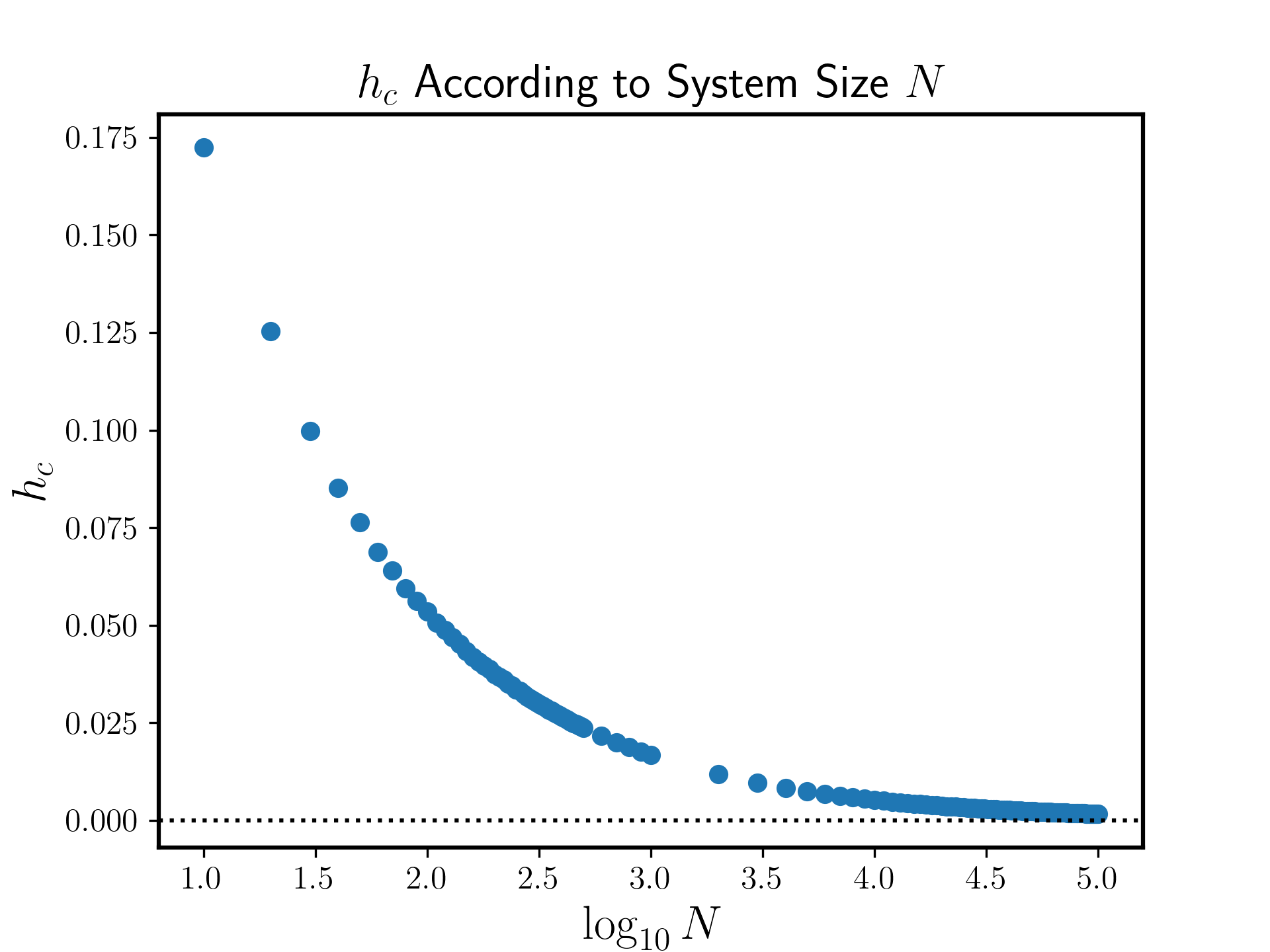}
    \caption{The critical values of the transverse field~(\ref{eq:hc}) $h_c$, with respect to the system size $N$ for $N\leq 10^5$ is illustrated. The dotted black line marks $h_c = 0$, as converges to it in thermodynamic limit, $N\to\infty$.}
    \label{fig:scal}
\end{figure}

We have shown that $h_c$ converges to $0$ in the thermodynamic limit in Fig.~\ref{fig:scal}. However, it should be noted that $h_c$ is significantly non-zero for finite system sizes. This implies that as the system size increases, the quantum effects will not survive at considering temperature regimes, and then it will undergo an order-disorder phase transition, consistent with the EA classical spin glasses~\cite{Rieger2005,TillHuse95}. 

Reducing disorder in the system can be seen as achieving a uniform transverse field distribution. Even if a uniform $h$ field is achieved, the system still contains randomness due to disorder in the $J_i$s. This leads to the question of whether randomly distributed interactions in spin-glass systems, which can have locally ordered spin regions, might influence the thermodynamics of the entire system. To better understand the impact of disordered interactions in spin-glass systems, we will discuss related phenomena, such as Griffiths phases~\cite{PhysRevLett.23.17} occur near the critical point.

\begin{itemize}
    \item The ``weakly disordered'' Griffiths phase occurs when $\max(\{J_i\}) > \min(\{h_i\}) = h$ but $\Delta_h > \Delta_c$,
    \item The ``strongly disordered'' Griffiths phase occurs when $ \max(\{J_i\}) < \min (\{h_i\}) = h$ and $\Delta_h > \Delta_c$,
    \item The ``weakly ordered'' Griffiths phase occurs when $\min(\{J_i\}) < \max(\{h_i\}) = h$ but $\Delta_h < \Delta_c$,
    \item The ``strongly ordered'' Griffiths phase occurs when $\min(\{J_i\}) > \max(\{h_i\}) = h$ and $\Delta_h < \Delta_c$.    
\end{itemize}
The points where the system enters these phases are not real phase transition points, but they nevertheless mark an important change in system behavior~\cite{Fisher1995}. However, disorder generally has stronger effects on quantum phase transitions than on classical transitions~\cite{TVojta2013}. We will primarily investigate the thermodynamic properties of the spin-glass system through quantum Otto heat engines and refrigerators.

\section{\label{sec:QOtto} The Quantum Otto Engine}

\begin{figure*}
    \centering
    \includegraphics[scale = 0.3]{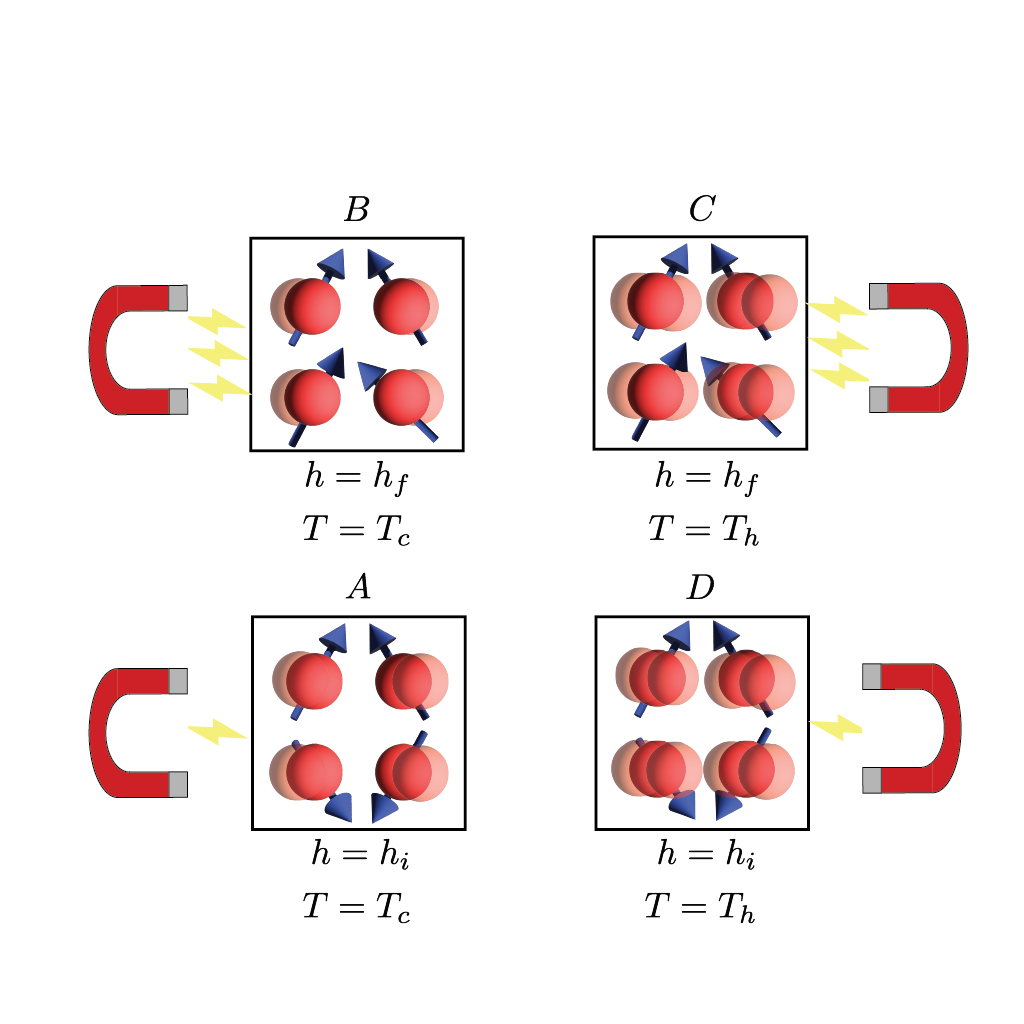}
    \caption{Schematic representation of a quantum Otto cycle composed of four strokes as adiabatic expansion ($A\to B$) where the system is decoupled from the baths while the magnetic field is transitioned from $h_i$ to $h_f$, isochoric hot thermalization ($B\to C$) where the system is coupled with the heat bath at temperature $T_h$ while the Hamiltonian $H(t)$ of the system is held constant, adiabatic compression ($C\to D$) where the system is decoupled from the baths while the magnetic field is transitioned from $h_f$ to $h_i$, and isochoric cold thermalization ($D \to A$) that the system is coupled with the cold bath at temperature $T_c$ while the Hamiltonian $H(t)$ of the system is held constant.}
    \label{fig:ottil}
\end{figure*}

The quantum Otto cycle (QOC) consists of four strokes. Two of these strokes are adiabatic while the other two are isochoric strokes. 

A quantum adiabatic stroke is a process in which no heat is exchanged while some work is performed. The energy levels of a Hamiltonian vary under such a stroke. In our QOE, the adiabatic stroke is performed via varying a magnetic field from a value $h_i$ to a value $h_f>h_i$, and vice versa.

In a quantum isochoric process, contact between a heat bath and the substance is established, causing a change in temperature. In such a process no work is done and the Hamiltonian eigenvalues are unchanged. In our QOE, our working substance, a quantum spin glass, is put in contact with baths at temperature $T_c$ and $T_h$, with $T_h > T_c$, respectively \cite{Quan_2007, Piccitto_2022}.

The strokes of the cycle are enumerated as follows:
\begin{enumerate}
    \item $A \rightarrow B$: Adiabatic expansion. The system is decoupled from the baths while the magnetic field is transitioned from $h_i$ to $h_f$.
    \item $B \rightarrow C$: Isochoric heating. The system is coupled with the heat bath at temperature $T_h$ while the Hamiltonian $H(t)$ of the system is held constant.
    \item $C \rightarrow D$: Adiabatic compression. The system is decoupled from the baths while the magnetic field is transitioned from $h_f$ to $h_i$.
    \item $D \rightarrow A$: Isochoric cooling. The system is coupled with the cold bath at temperature $T_c$ while the Hamiltonian $H(t)$ of the system is held constant.    
\end{enumerate}

\subsection{\label{subsec:regimes}Working Regimes of the QOE}

The four working regimes of the QOE, including the quantum refrigerator (R) and heat engine (HE), along with the relevant quantities. These quantities include work output and the quantity to analyze the enhancing effects of quantum spin glasses, i.e. thermodynamic performance.
We are going to deal with three relevant quantities from the Otto Engine:
\begin{itemize}
    \item $Q_c$: The heat exchanged with the cold reservoir
    \item $Q_h$: The heat exchanged with the hot reservoir
    \item $W$: The total work done during a cycle is equal to the sum of heat exchanged with the baths ($W = Q_c+Q_h$)
\end{itemize}
Combining the work equation with the celebrated Clausius inequality of thermodynamics, we see that only four working regimes are allowed~\cite{Solfanelli_2020}. These are called heat engine (E), refrigerator (R), heater (H), and accelerator (A) regimes.
\begin{table}[h!]
  \centering
  \label{tab:table1}
  \caption{Comparison of different working regimes in terms of work output and heat exchanged.}
  \begin{tabular}{@{}p{3cm}p{2em}p{2em}p{2em}@{}}
    \toprule
    Working Regime & \( W \) & \( Q_c \) & \( Q_h \) \\
    \midrule
    Heat Engine (E) & \( > 0 \) & \( > 0 \) & \( < 0 \) \\
    Refrigerator (R) & \( < 0 \) & \( > 0 \) & \( < 0 \) \\
    Heater (H) & \( < 0 \) & \( < 0 \) & \( < 0 \) \\
    Accelerator (A) & \( < 0 \) & \( < 0 \) & \( > 0 \) \\
    \bottomrule
  \end{tabular}
\end{table}

By convention, note that $Q_{c (h)}>0$ implies the engine absorbs heat from the cold (hot) reservoir, and for a complete cycle $Q_{c(h)} = -Q_{h(c)}$.

Let's consider a system in the Otto cycle, where the state is at point $B$ before thermalization and at point $C$ after thermalization, with the time-dependent Hamiltonian being $H(\tau)$ at time $t = \tau$. Since, by definition, no work is done during a quantum isochoric process, the change in internal energy depends only on the heat exchanged. Thus,
\begin{equation}
  Q_c = \Delta U = \langle H(\tau) \rangle_{\rho(C)}-\langle H(\tau) \rangle_{\rho(B)},  
\end{equation}
where $\rho(P)$ denotes the density matrix of the system at any point $P={A,B,C,D}$.

To find the work output $W$, note that the total change in internal energy over one complete cycle is zero. Thus, $\Delta U_{\text{tot}} = Q_{\text{tot}} - W_{\text{tot}}=0$. This leads to $W_{\text{tot}} = Q_{\text{tot}} = Q_c + Q_h$. For further information on calculating Hamiltonian expectation values, refer to Appendix~\ref{app:exp}. Keep in mind that our calculations are based on the ideal application of the adiabatic theorem \cite{PhysRevLett.131.060602}.

The engine efficiency is defined as
\begin{equation}
 \eta = W/Q_h,   
 \label{eq:eff}
\end{equation}
in terms of work output $W$ and heat absorbed from the hot reservoir, $Q_h$. The Carnot efficiency is given by $\eta_C = 1-T_c/T_h$, where $T_c$ and $T_h$ represent the cold and hot reservoir temperatures, respectively.

In quantum heat engines, increasing the system size typically leads to a rise in work output, but it often results in decreased efficiency. From this perspective, thermodynamic performance, $\Pi$, is relevant because it accounts for both factors. The main quantity that we will consider is the ``thermodynamic performance'',
\begin{equation}
   \Pi = \frac{W}{\delta\eta}, 
   \label{eq:Pi}
\end{equation}
where $\delta\eta = \eta_C-\eta$ is the measure of how close the engine's efficiency is to the Carnot efficiency. Superlinear scaling in $\Pi$ would suggest that the increase in work output outweighs the decrease in efficiency as the system grows, indicating improved performance \cite{Campisi2016vu}. The present study aims to demonstrate that superlinear scaling occurs due to quantum criticality in quantum spin glass systems.

\section{\label{sec:res} Results}

In this section, we will present our results regarding the proposed QOE. Unless stated otherwise, the fixed quench is applied
\begin{equation}
    \delta h = h_f - h_i = 0.5,
\end{equation}
All our data are averaged over $512$ realizations of the independent $J_i$ Gaussian distribution with zero mean and the variance $1/N$. 
\begin{figure}
    \centering
    \includegraphics[scale = 0.5]{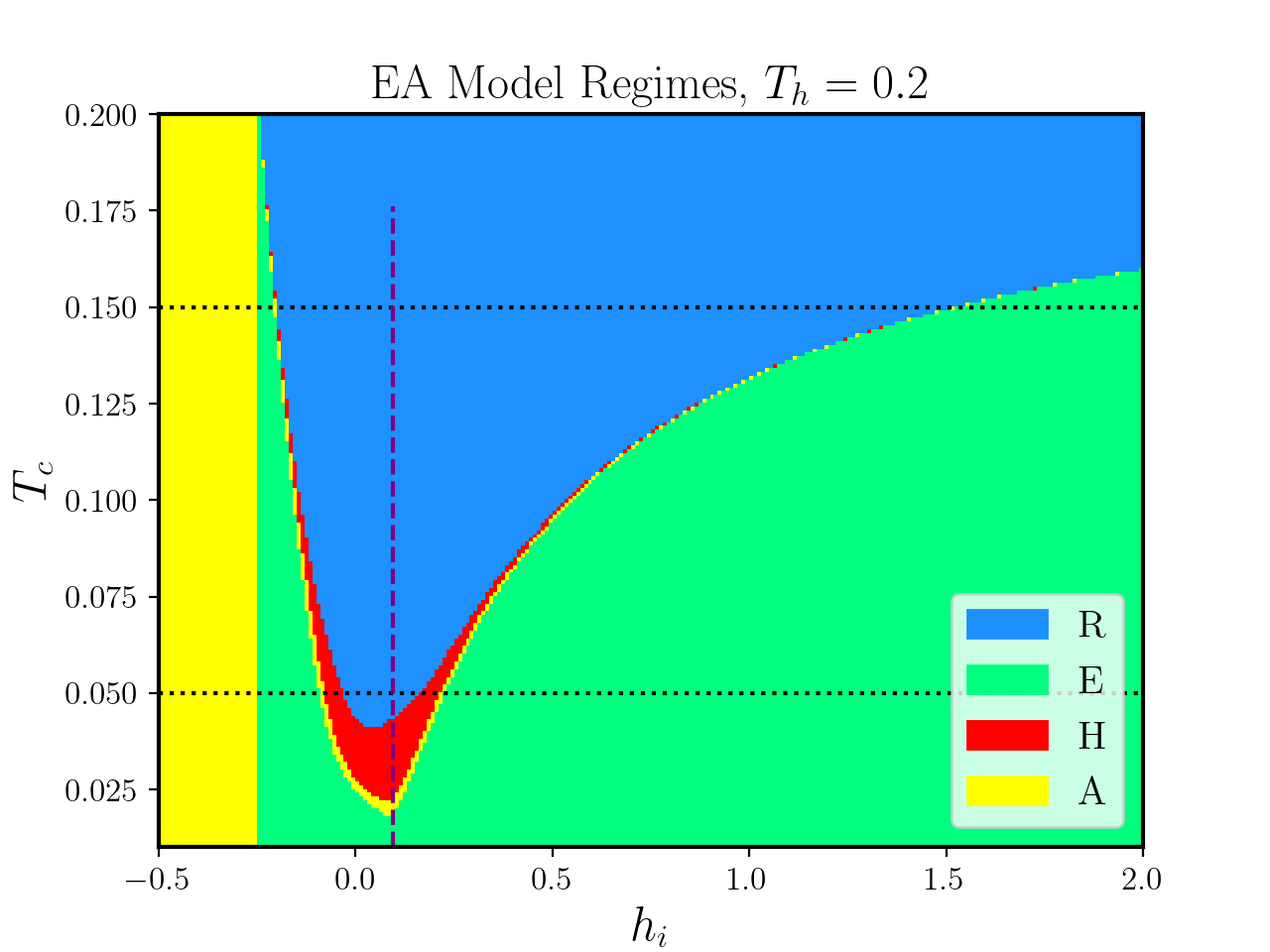}
    \includegraphics[scale = 0.5]{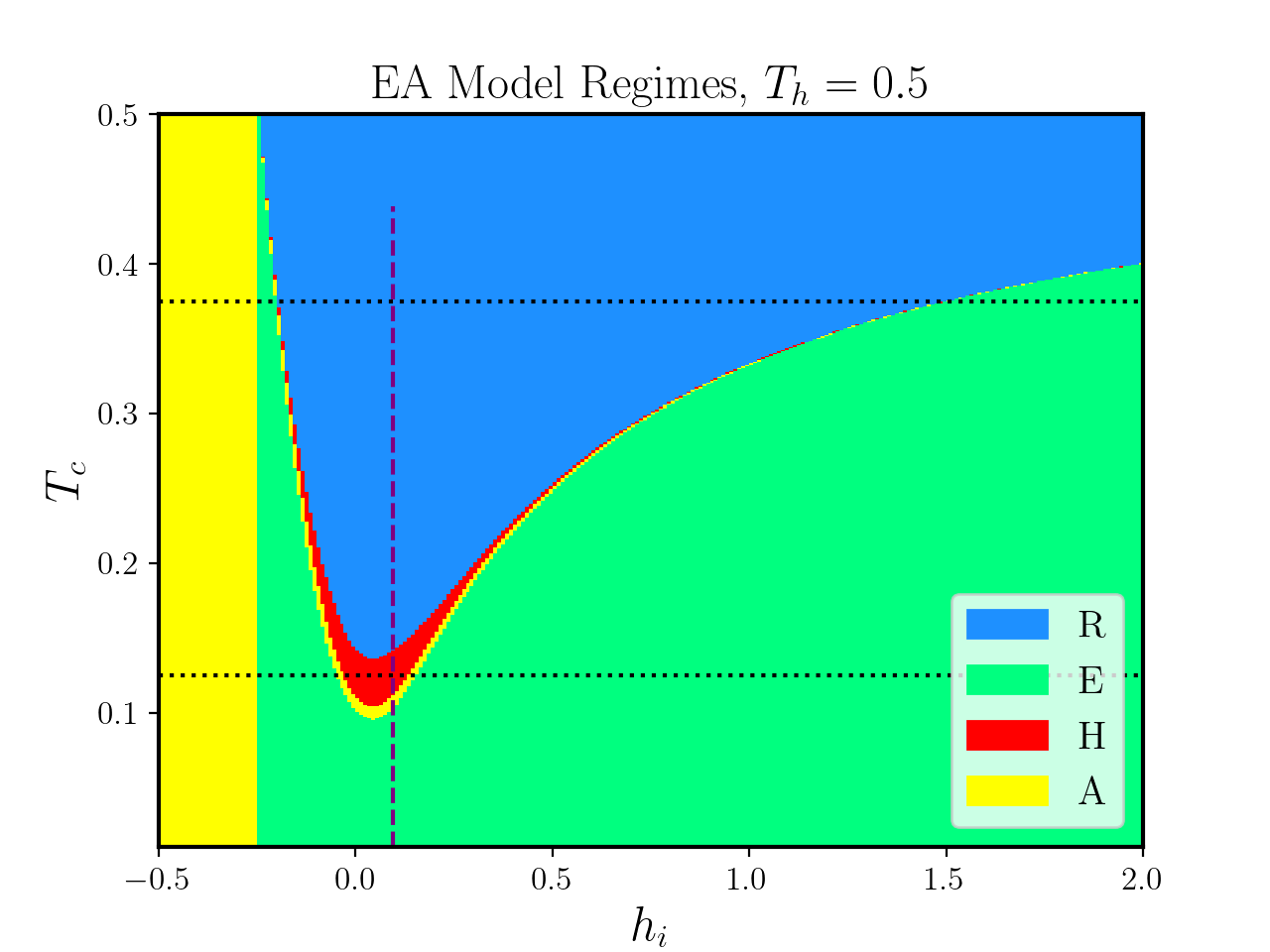}
    \caption{Working Regimes of the Otto Engine (EA) for $T_h=0.2,0.5$. The purple lines denote $T_c = T_h/4$ and the  black lines denote $T_c = 3T_h/4$, see Table I.}
    \label{fig:regs}
\end{figure}

We present the operating regimes of our QOE with a continuous range for $0\leq T_{c}\leq T_h$ and $-0.5\leq h_i\leq 2$, while varying $T_h=0.2$ and $0.5$. A detailed explanation for these choices will be provided as follows. However, it should be noted that once we understand the behavior of the working regimes depending on the related parameters, it can certainly be tuned to operate in the specific mode required for the Otto engine.

In Figure~\ref{fig:regs}, it is evident that as the temperature of the cold reservoir gets closer to that of the hot reservoir, it becomes more likely to achieve a refrigerator regime. 
\begin{figure*}[!ht]
    \centering
    \includegraphics[scale = 0.5]{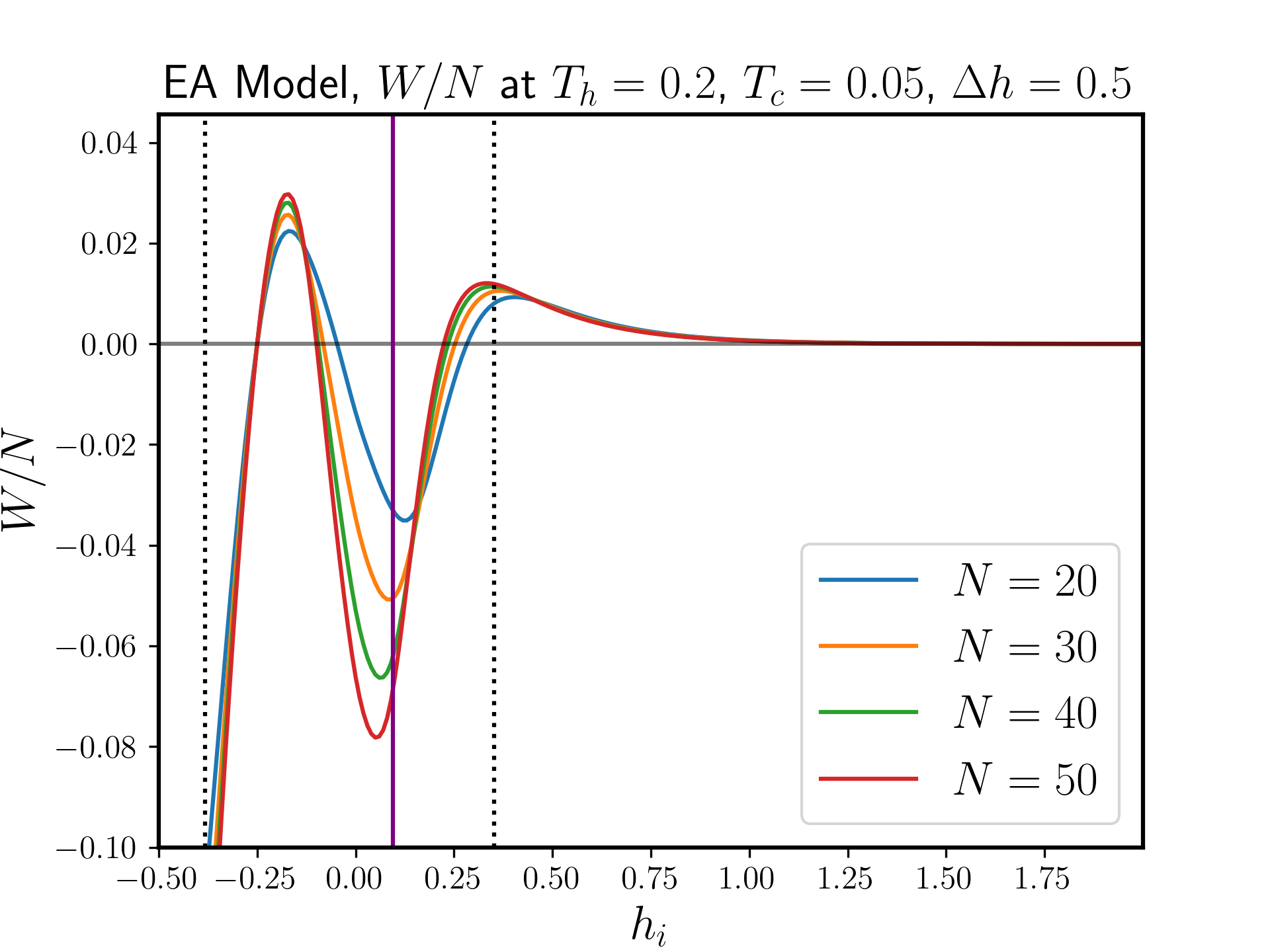}
    \includegraphics[scale = 0.5]{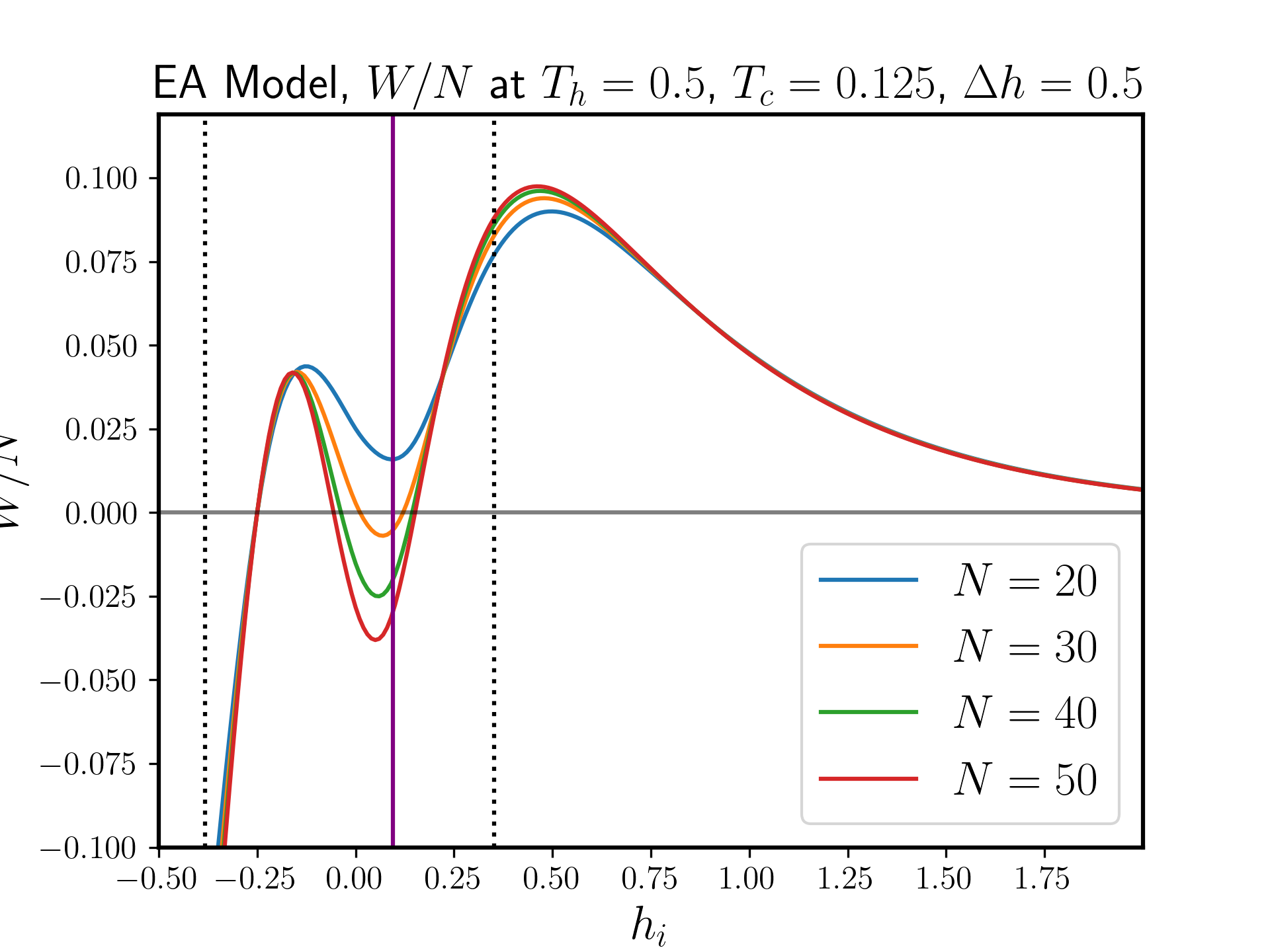}\\
    \includegraphics[scale = 0.5]{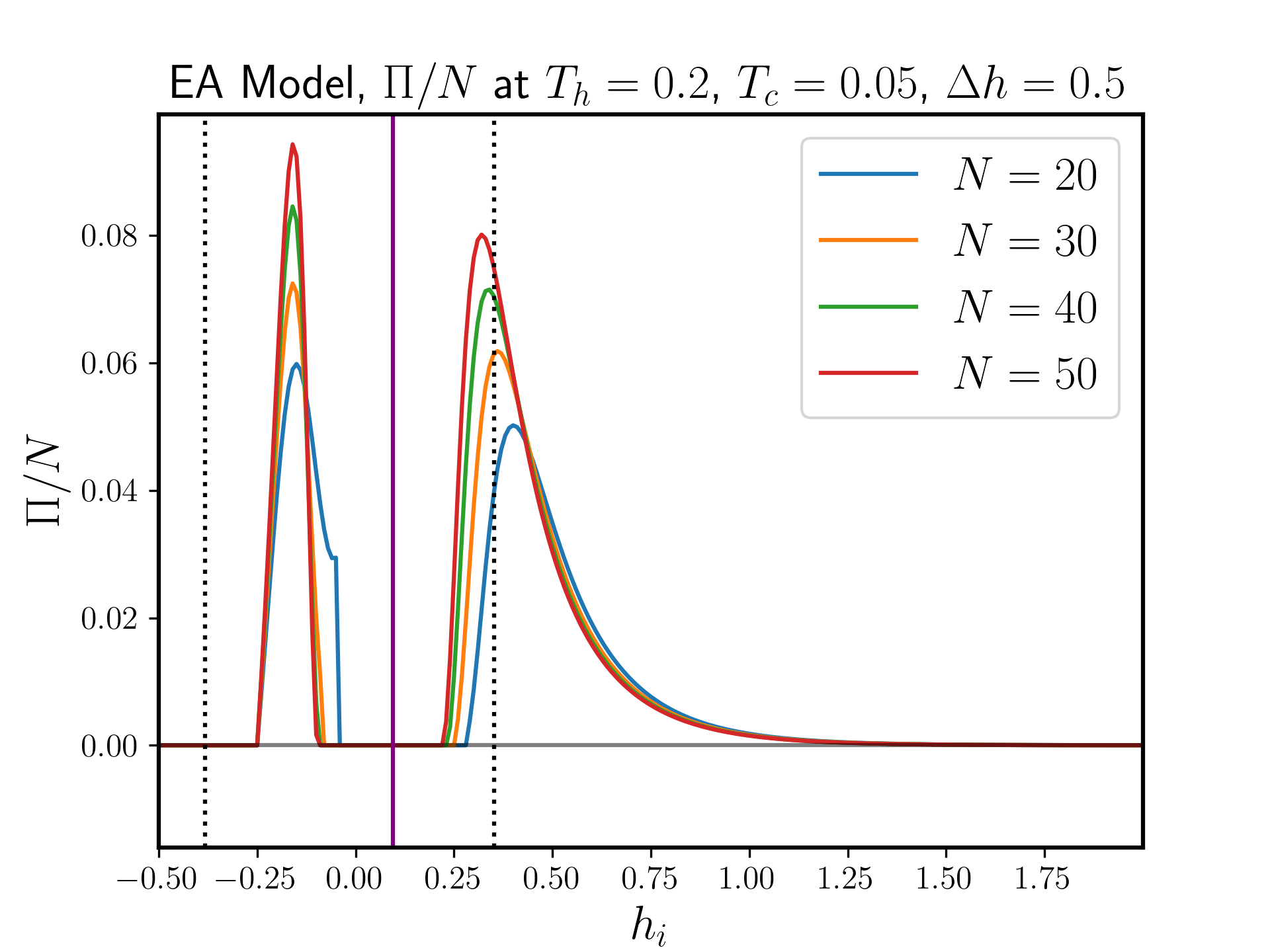}
    \includegraphics[scale = 0.5]{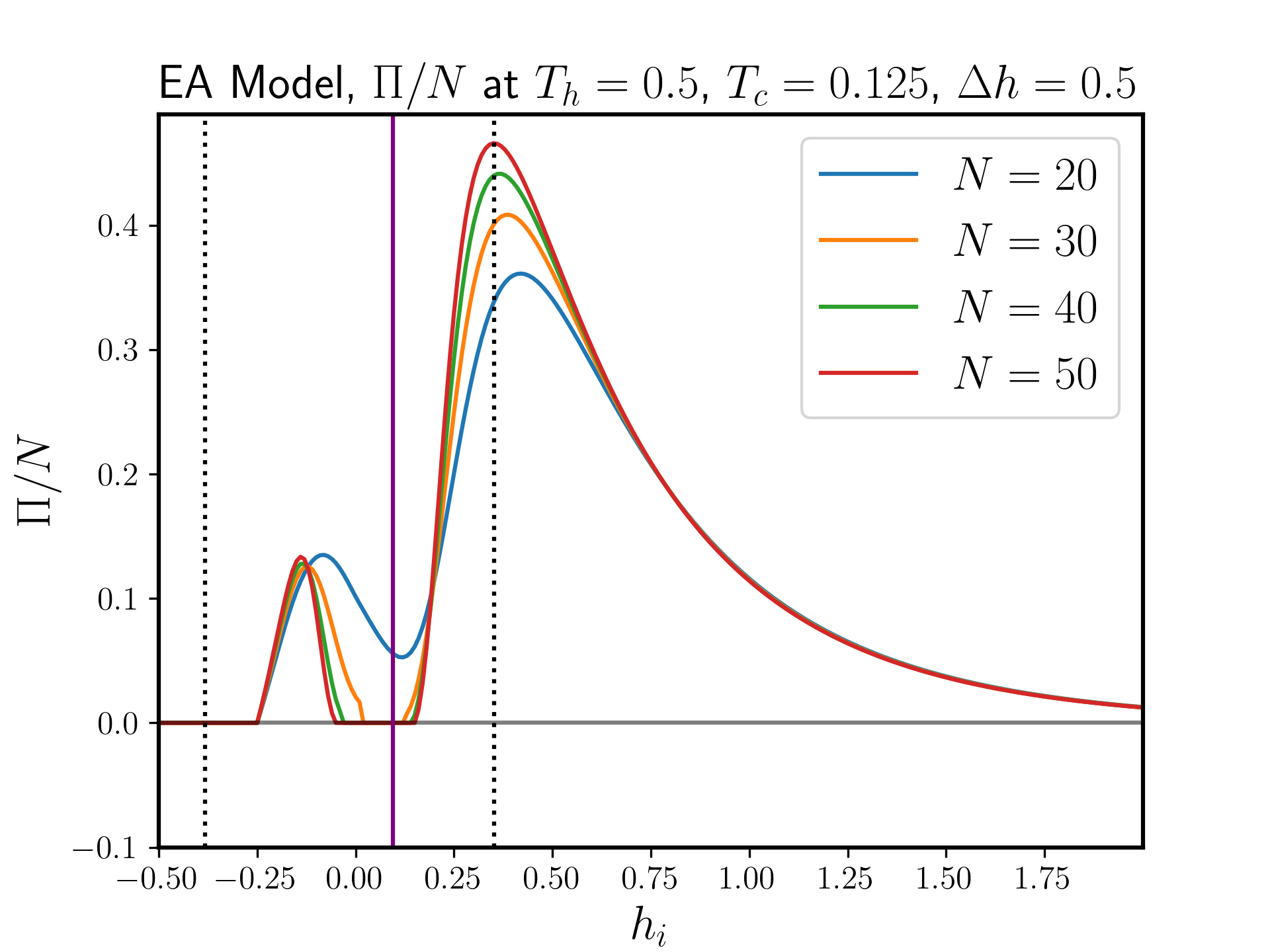}    
    \caption{Top line: Comparison of Work per Spin, $W/N$ for $T_h = 0.2,0.5$, $T_c = T_h/4$, with $\delta h = 0.5$ held constant. Number of realizations is 512. The solid purple line marks the phase transition point where $\Delta_h = \Delta_c$ and the dashed black lines mark the ends of the Griffiths phases. Bottom Line: Comparison of Thermodynamic Performance per Spin, $\Pi/N$ for $T_h = 0.2,0.5$, $T_c = T_h/4$, with $\delta h = 0.5$ held constant. Number of realizations is 512. The solid purple line marks the phase transition point where $\Delta_h = \Delta_c$ and the dashed black lines mark the ends of the Griffiths phases. The values of $\Pi/N$ where $W/N<0$ were clipped to $0$ to avoid any unwanted peaks as our definition of $\Pi$ assumes $W/N>0$.}
    \label{fig:wps}
\end{figure*}

Conversely, as the difference between $T_h$ and $T_c$ increases, a heat engine regime becomes more probable. Furthermore, there is a notable region between the boundaries of the refrigerator (R) and heat engine (E) regimes, encompassing other regimes such as heater (H) and absorption (A). As we will observe in Section \ref{subsec:heng}, the minimum point of this region consistently lies between the two peaks in work output and thermodynamic performance, as shown in Figures~\ref{fig:wps}, separating them.

We can identify a trend by looking at the sequence of regimes from high to low $T_c$. Initially, as $T_c$ decreases, $Q_c$ shifts from positive to negative. With further reduction in $T_c$, $Q_h$ also changes sign from positive to negative. Finally, as $T_c$ continues to decrease, the signs of both these quantities invert, resulting in positive work output. This is convenient with the $2$nd law of Thermodynamics.

Another intriguing observation is the predictable shift in the boundaries of the E and R regimes as $T_h$ increases. These changes, which we will delve into in the following subsections, highlight the predictable nature of thermodynamics and the control we can exert over these regimes.

\subsection{\label{subsec:heng} Heat Engine Regime}
This section will examine the work output $W$, and the thermodynamic performance, $\Pi$, of our QOE operating as a heat engine.

Let's consider the fixed $T_c$ temperature that will allow the Otto engine to operate in the heat engine regime as $T_h/4$ for two cases, where $T_h=0.2$ and $0.5$, see Fig.~\ref{fig:wps}. 

A notable observation is that $W/N$ exhibits a double-peaked structure with respect to $-0.5\leq h_i\leq 1.75$. The reason for selecting $T_h = 0.2$ and $0.5$ becomes clear: For $T_h = 0.2$, the first peak yields a higher work output than the second one, whereas, for $T_h = 0.5$, the second peak provides a higher work output. This distinction between peaks is an important theme throughout the paper.

\begin{figure}
    \centering
    \includegraphics[scale = 0.5]{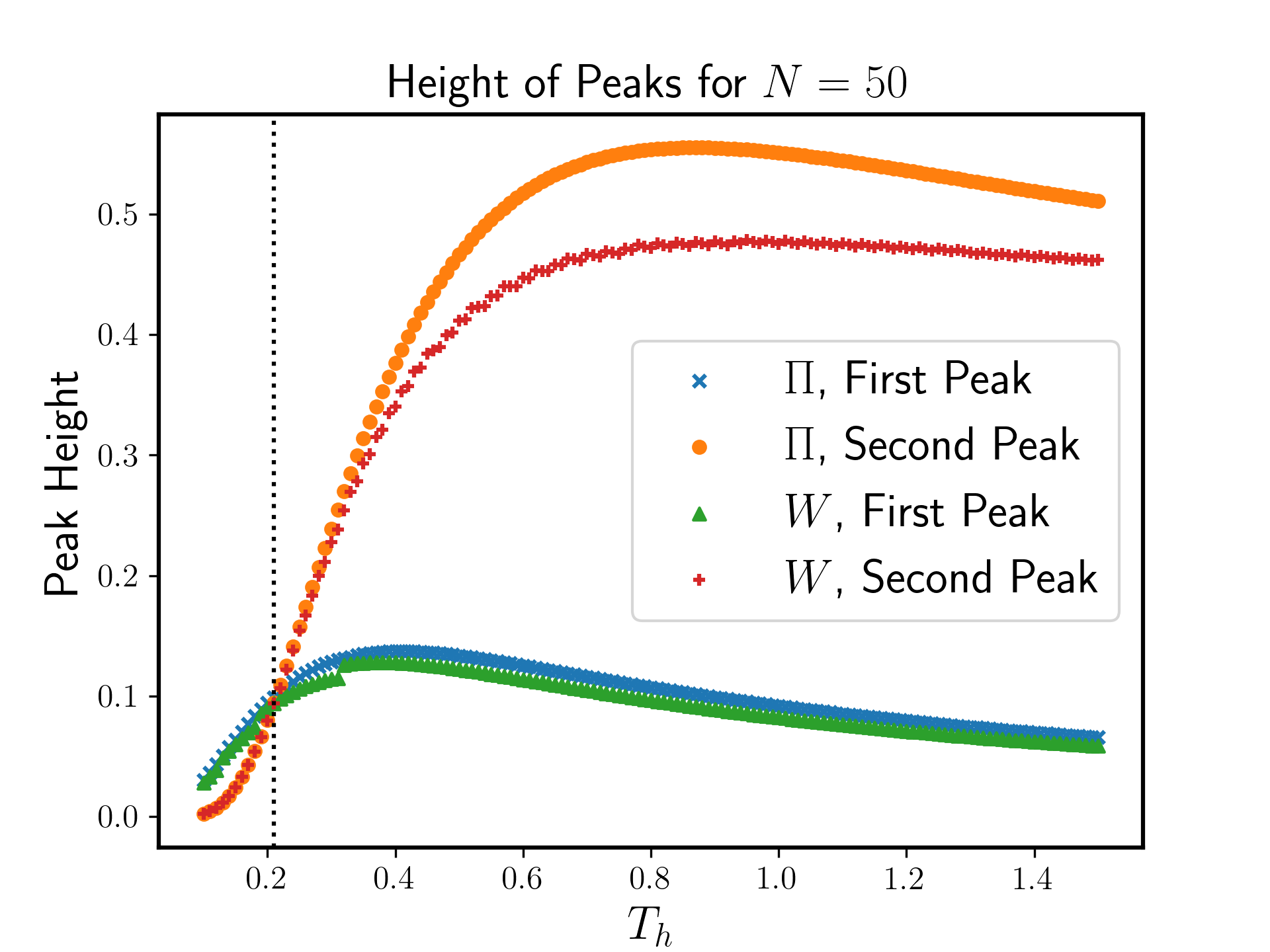}
    \caption{Comparison of the height of the two peaks. The dashed black line marks the intersection at $T_h \approx 0.29$. From now on, including this figure, the orange marks denote the second peak while the blue marks denote the first. Peak height for thermodynamic performance per spin, $\Pi/N$.}
    \label{fig:pheight}
\end{figure}

In Fig.~\ref{fig:pheight}, it is seen that a similar double-peaked structure is observed for thermodynamic performance. As most behavior of $\Pi/N$ is similar with $W/N$, we'll only consider the two significantly different behavior: Peak height and scaling of the classical peak.

To determine the dominant parameter scaling with respect to work output and thermodynamic performance, we demonstrate how the positions of the maxima, calculated for a system with $N=50$ spins, vary with $T_h$ over the range $0.05 \leq T_h \leq 1.5$.

We observed that the temperature at which the dominant peak shifts is approximately $T_h \approx 0.29$, as depicted in Fig.~\ref{fig:pheight} at the crossing point. For $T_h < 0.29$, the first peak is dominant, but the second peak becomes dominant when the temperature exceeds this value.

As depicted in Fig.~\ref{fig:wps}, the critical point $h_c$ separates the two peaks. Consistent with the hypothesis proposed in \cite{Campisi2016vu}, the first peak corresponds to a quench that crosses the critical point. This is because $h_i < h_c < h_f = h_i + \Delta h = h_i+0.5$ where $h_i$ represents the peak position on the graph. Additionally, the first peak typically resides within the weakly ordered Griffiths phase, while the second peak exhibits behavior largely independent of this phase. Following the second peak, we notice an exponential tail, indicating the formation of larger rare clusters within the system~\cite{TV2003}. This observation suggests that a regime characterized by this exponential tail has replaced the weakly disordered Griffiths phase~\cite{TV2004,TV2008}.

Given these observations and considering that lower temperatures tend to favor the first peak, we refer to the first peak as the ``quantum peak'' and the second one as the ``classical peak''. 

\subsection{Superlinear Scaling of Ising Spin-glass Quantum Otto Heat-engine}

\begin{figure}[h!]
    \centering
    \includegraphics[scale = 0.5]{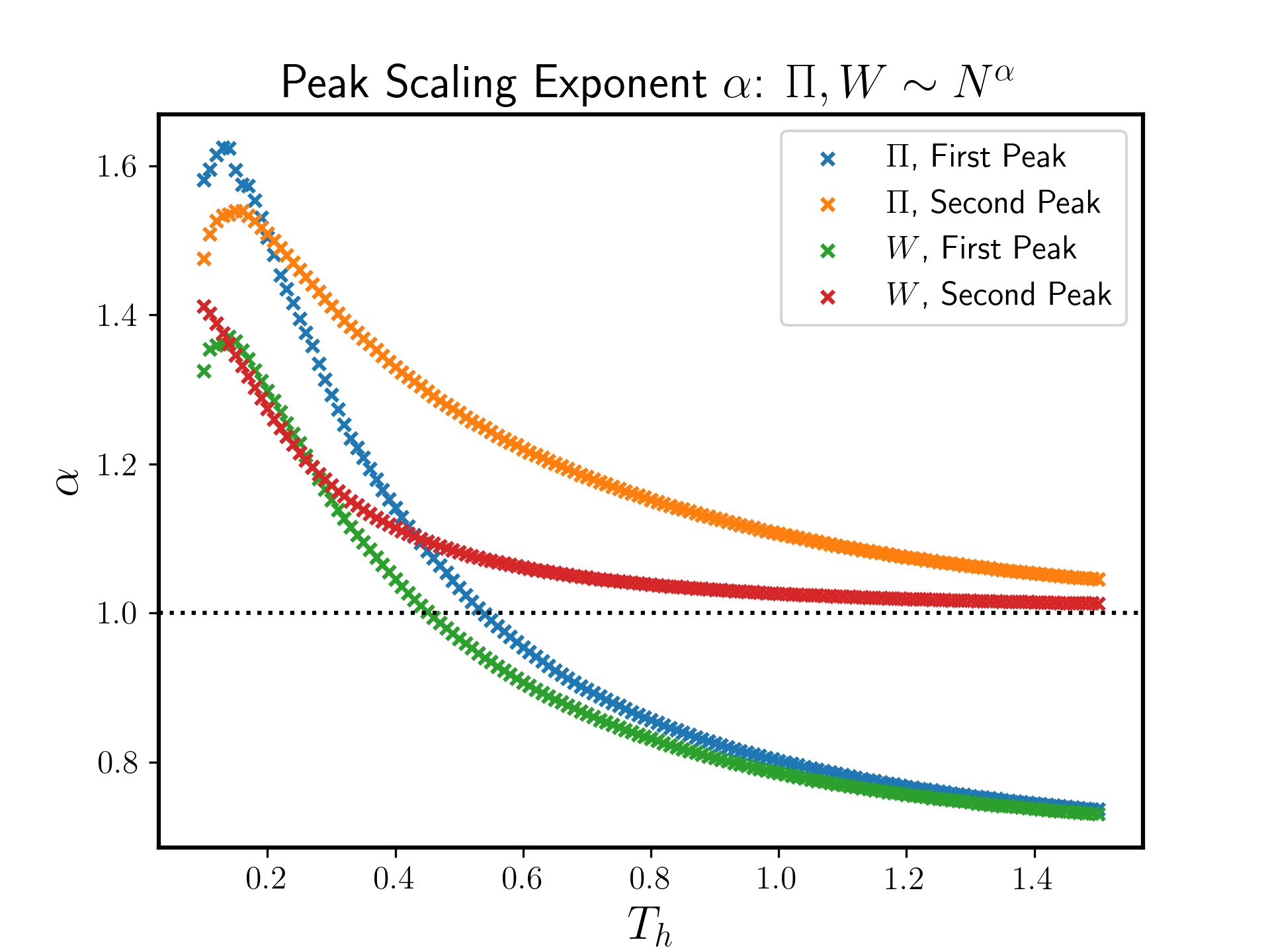}
    \caption{The scaling behavior for work output, $W$, and thermodynamic performance, $\Pi$, with respect to the hot bath temperature $T_h$ is given for both peaks by the scaling exponent $\alpha$ such that the relevant quantity,$Q$, obeys the scaling $Q\sim N^\alpha$. The dashed black line denotes $\alpha = 1$ which corresponds to the linear scaling with system size. }
    \label{fig:pscal}
\end{figure}

In Fig.~\ref{fig:pscal}, we show that the exponents with which the peaks scale with respect to the system size, that is, $\alpha$ for $W \sim N^\alpha$, hold true. We achieve the numerical values for $\alpha$ via fitting a function of form $y = bN^\alpha$ for the values of the relevant quantities at the peaks.

It is important to note that the scaling of the quantum peak decreases from superlinear to sublinear with $T_h$, while the scaling of the classical peak converges to $1$ with increased temperature. This is a significant deviation from the pure model with uniform $J$ in many ways. Following the calculations in~\cite{Piccitto_2022}, we observed in our numerical simulations that the quantum peak dissolves, and the classical peak starts from sublinear and converges to linear scaling with increased temperature. Thus, we conclude that the randomness and frustration incorporated into the system affect the work output quantum peak anomalously, which may lead to enhancement in low temperatures, along with inverting the scaling of the classical peak, albeit leaving the value it converges to unchanged at $\alpha = 1$.

We are going to use the definition of thermodynamic performance, $\Pi$ in~(\ref{eq:Pi}) and search for the enhancement proposed in \cite{Campisi2016vu}. However, we must note that the 1D Edwards-Anderson model in a transverse field exhibits anomalous scaling behavior when compared with other second-order quantum phase transitions and that the dimension $d=1$ is below the lower critical dimension for such a system~\cite{Fisher1995,Young_1997}. Thus, our scaling isn't expected to strictly obey the polynomial scaling in \cite{Campisi2016vu}. 
\begin{figure*}[ht]
    \centering
    \includegraphics[scale = 0.52]{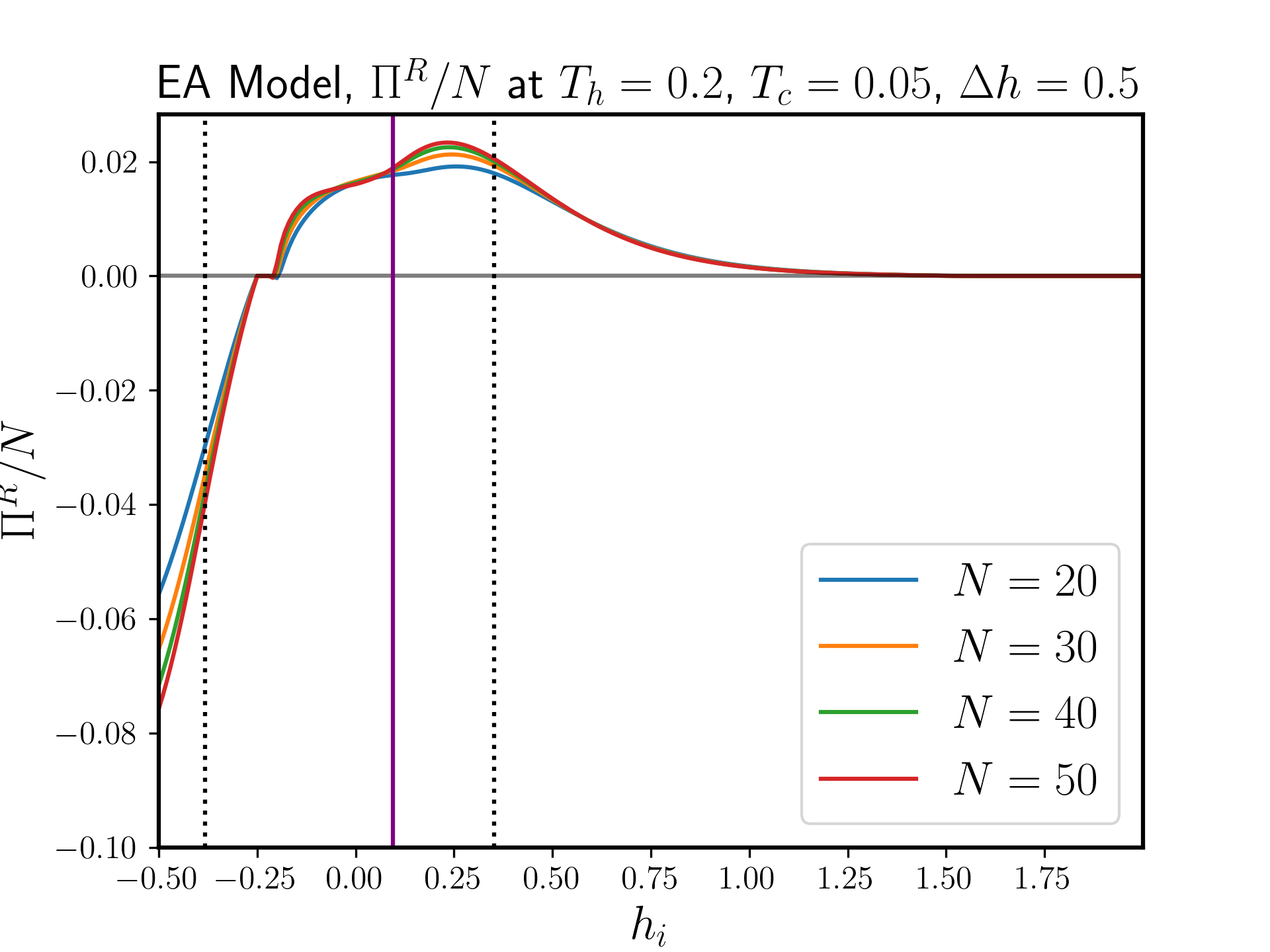}
    \includegraphics[scale = 0.52]{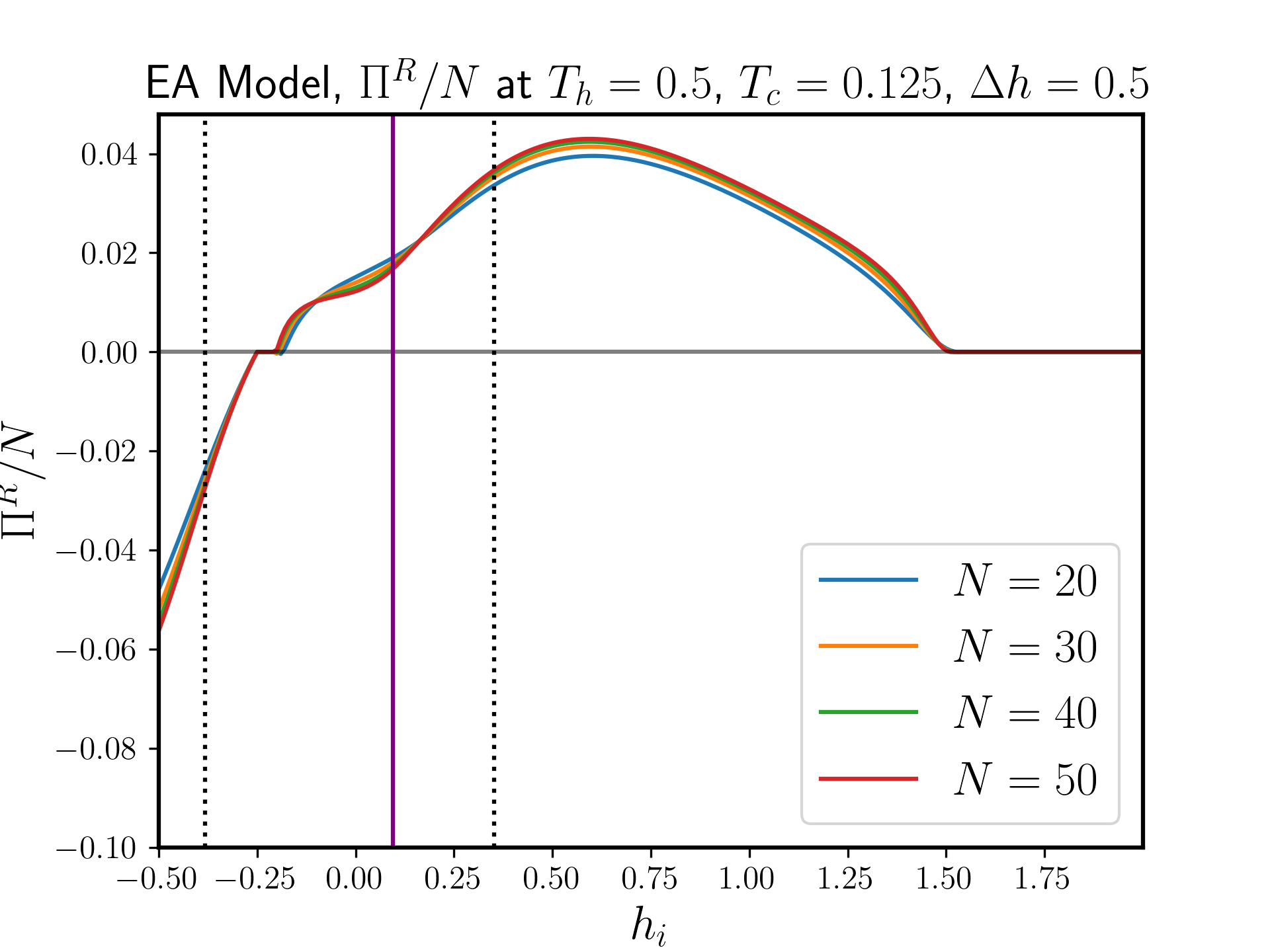}
    \caption{Comparison of Refrigerator Performance per Spin, $\Pi_R/N$ for $T_h = 0.2,0.5$, $T_c = 3T_h/4$, with $\delta h = 0.5$ held constant. Number of realizations is 512. $\Pi^R$ was clipped to $0$ where $W>0$.}
    \label{fig:pirps}
\end{figure*}

As seen in Fig.~\ref{fig:pheight}, in thermodynamic performance, the peak height of the classical one also reaches a maximum rather than approaching asymptotic convergence, as observed for work output's classical peak. Moreover, in~Fig.~\ref{fig:pscal}, the convergence of the classical peak to $\alpha=1$ is slower along with the scaling drop in the quantum peak compared to work output. 

Thus, we conclude that we see a greater enhancement in $\Pi$ compared to $W$, signaling a significant enhancement in the heat engine efficiency, $\eta$.

Overall, we observe significant enhancement possibilities in both work output, $W$, and thermodynamic performance, $\Pi$, due to the disorder and frustration involved in the system.

\subsection{\label{subsec:ref} Refrigerator Efficiency}

In this subsection, we will examine the performance of our QOC as a quantum refrigerator. To achieve this, we must change our definition of performance.

Denote the refrigerator performance by $\Pi_R$. Similar to our definition for $\Pi$, we define 
\begin{equation}
    \Pi_R = Q_c/\delta\eta_R
    \label{eq:PiR}
\end{equation}
where $\delta\eta_R$ is defined analogously as $\delta\eta_R = \eta_{\text{COP}}-\eta_R$. Here, $\eta_{\text{COP}} = T_c/(T_h-T_c)$ is the Carnot coefficient of performance, and
\begin{equation}
   \eta_R = Q_c/|W| = -Q_c/W
    \label{eq:etaR}
\end{equation}
is the refrigerator efficiency.

Following these definitions, we obtained Fig. \ref{fig:pirps}, where $T_h = 0.2,0.5$ as before and $T_c = 3T_h/4$. As seen in Fig. \ref{fig:regs}, the engines are in the refrigerator regime for most $h_i$ for such a choice of $T_c$.

From Fig. \ref{fig:pirps}, we observe arising structure with a slope followed by a shoulder and then a peak in which the shoulder dissolves quickly and the peak widens as temperature increases. We will focus on the peak. However, it is important to note that the curves for different system sizes overlap at a point that corresponds to direct quenching over the quantum critical point. The peak appears to be independent of the quantum critical point and dependent on temperature. The two relevant quantities we will examine are the quench midpoint of the peak and the scaling factor $\alpha$.
\begin{figure}
    \centering
    \includegraphics[scale = 0.5]{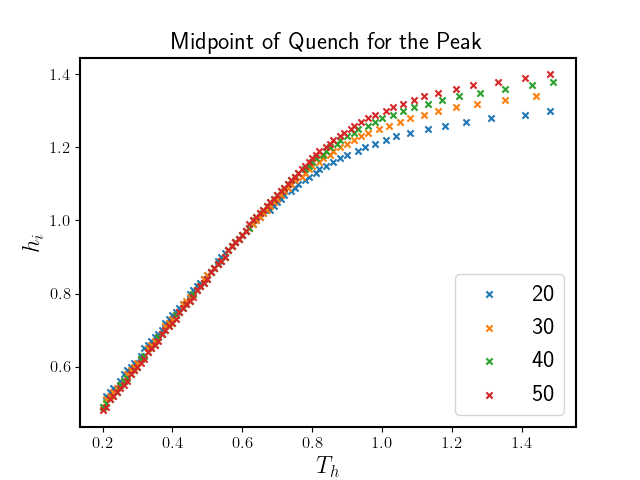}
    \caption{Quench midpoint for refrigerator performance, $\Pi^R$.The point where peaks diverge is roughly around $T_h\approx0.7$ while $h_i\approx0.9$.}
    \label{fig:pirmqp}
\end{figure}

\begin{figure}
    \centering
    \includegraphics[scale = 0.5]{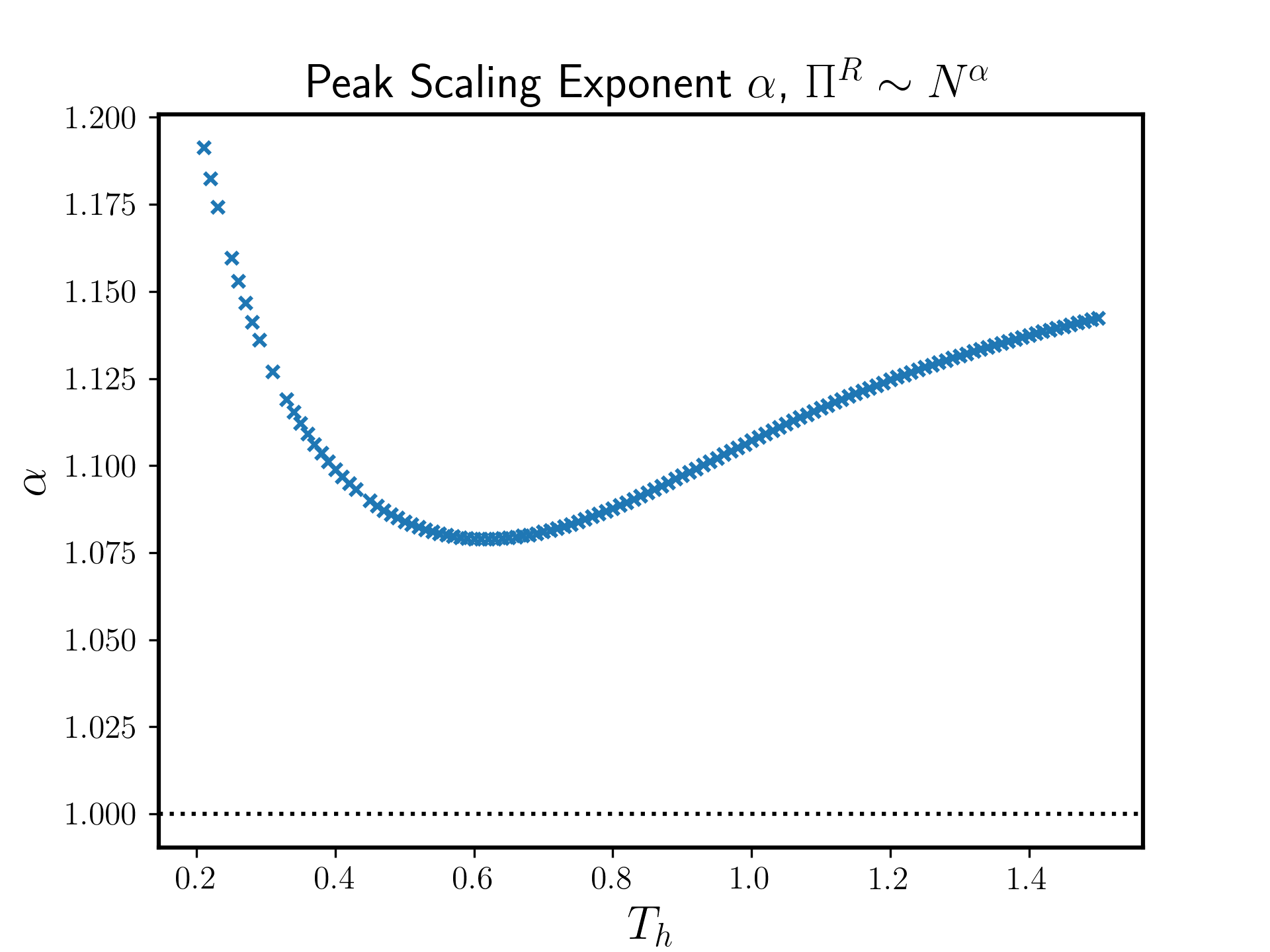}
    \caption{The scaling behavior for the refrigerator performance, $\Pi^R$, with respect to the hot bath temperature $T_h$ is given for the second peak by the scaling exponent $\alpha$ for refrigerator performance, $\Pi^R \sim N^\alpha$. The dashed black line denotes $\alpha = 1$ which corresponds to the linear scaling with system size.}
    \label{fig:pirscalh}
\end{figure}

An important observation from Fig. \ref{fig:pirmqp} is that the location follows a linear shape at low temperatures; however, the curves for system sizes start to differentiate and converge to distinct values at higher temperatures. This divergence starts roughly around $T_h \approx 0.7$. From Fig. \ref{fig:pirscalh}, we infer that unlike in the heat engine regime, the scaling with system size converges to an $\alpha > 1$, to roughly around $\alpha \approx 1.1$, giving genuine superlinear scaling which does not disappear with temperature for $T_h \leq 1.5$. Thus, by extrapolating our results, we conclude that for high temperatures, the refrigerator performance peak is localized and shows superlinear scaling with a constant exponent, surpassing the results of the pure model in~\cite{Piccitto_2022}. For low temperatures, we obtain even higher scaling which falls rapidly to a minimum and rises, along with the location of the peak changing rapidly.

\section{Conclusions}
\label{sec:conc}

In conclusion, we studied the 1D Ising spin-glass model in a transverse field as a working medium in the quantum Otto cycle. We investigated both heat engine and refrigerator regimes in relatively low and high temperature scales by tuning the hot and cold bath temperatures. First, we mapped our model to the free fermions model via the Jordan-Wigner transformation and then we calculated the relevant quantities for $N=20,30,40,50$ spins. 

The primary focus of our investigation is the genuine superlinear scaling in the performance of work output and engine performance in both the heat engine and refrigerator operating regimes. In the case of the heat engine regime, we observe a double-peaked structure, whereas, for the refrigerator, we obtain a shoulder and then single-peaked structure similar to the Ising model~\cite{Piccitto_2022}. In the case of the double-peaked work output, we infer that the first peak is governed by quantum effects, while the second peak is influenced by classical effects, as observed through their behaviors at increasing temperatures.

The ``quantum'' peak directly corresponds to a quench over the quantum critical point, $h_c$, and that its location converges within the Griffiths phases\cite{RiegerYoung1996,Rieger,Rieger2005} while the ``classical'' peak is independent of the quantum critical point and its location increases linearly after a certain temperature, $T_h \approx 0.29$. This temperature also corresponds to the change in the domination of the peaks, as for $T_h < 0.29$, the quantum peak outputs higher work than the classical peak when for $T_h>0.29$, the classical peak corresponds to higher work output.

We also see that the scaling exponent $\alpha$ of the quantum peak goes from $\alpha >1$ to $\alpha <1$ rapidly, slowing its decrease down in higher temperatures. Thus, we find it possible to obtain superlinear scaling for lower magnetic field strengths only for lower temperatures. For the classical peak scaling, we see that scaling starts superlinear and converges to linear with increasing temperature. Thus, we conclude that genuine superlinear scaling from a quantum heat engine with a 1D chain quantum Ising spin glass, specifically the simplest spin-glass model known as the Edwards-Anderson model in a transverse field, is achievable for lower temperatures.

For the quantum refrigerator, we encounter a shoulder and following a peak structure with the shoulder dissolving and the second peak widening with increased temperature. This peak also behaves independently of the quantum critical point and the Griffiths phases; however, unlike the classical peak in the quantum heat engine, its location converges to a number greater than one in higher temperatures. Additionally, this peak exhibits genuine superlinear scaling even at higher temperatures, with $\alpha$ converging to a constant value greater than $1$ as the temperature increases. 

Hence, we conclude that it is possible to obtain genuine superlinear scaling from a quantum refrigerator with a $1D$ quantum spin-glass model in a transverse field for both lower and higher temperatures, significantly surpassing the results of the pure Ising model, i.e., without randomly distributed disorder,~\cite{Piccitto_2022}. 

Our study is the first to utilize quantum spin glasses as a fuel source in quantum heat engines, where we observed phase transitions through work output and obtained superlinear scaling at the critical point. The engine efficiency proved to be higher than that of trivial spin systems, with even greater performance in the refrigerator regime. These findings are not only groundbreaking in the fields of quantum thermodynamics and heat engines, but they also contribute to a deeper understanding of the nature of spin glasses, providing new insights for future research.

\begin{acknowledgments}
This study was supported by the Scientific and Technological Research Council of Turkey (TUBITAK) under the Grant Number 120F100. The authors thank to TUBITAK for their supports.
\end{acknowledgments}
\appendix

\section{\label{sec:bog}The Bogoliubov Transform}
As the Hamiltonian matrix grows exponentially with the system size, we use a special diagonalization method called the Bogoliubov Transform to find another related matrix that is linear in system size \cite{mbeng2020quantum}.
First, we must apply the Jordan-Wigner Transform to our Pauli matrices to convert them into free fermions. We will follow \cite{mbeng2020quantum} and \cite{PhysRevB.53.8486} in this process.

The Pauli matrices transform as follows,
\begin{equation}\displaystyle \sigma_j^x = \bigg[\Pi_{i=1}^j(c_i^\dag+c_i)(c_i^\dag-c_i)\bigg](c_j^\dag+c_j) \end{equation}
\begin{equation} \sigma_j^y = i\bigg[\Pi_{i=1}^j(c_i^\dag+c_i)(c_i^\dag-c_i)\bigg](c_j^\dag-c_j) \end{equation}
\begin{equation}\sigma_j^z = 1-2c_j^\dag c_j \end{equation}
Then, the spin coupling terms $\sigma_i^x \sigma_{i+1}^x$ becomes,
\begin{equation}\sigma_i^x \sigma_{i+1}^x = c^\dag_ic^\dag_{i+1} + c^\dag_ic_{i+1} + h.c. \end{equation}
which can be verified using $1-2c_i^\dag c_i = (c_i^\dag+c_i)(c_i^\dag-c_i)$, $ (1-2c_i^\dag c_i)^2 = 1 $ and the commutation-anticommutation relationships.

Equipped with the Jordan-Wigner Transform, we write our Hamiltonian,
\begin{equation}
\begin{split}
    H = -\sum_{i=1}^{N}J_{i}(a_i^\dag a_{i+1} + a_i^\dag a_{i+1}^\dag + \text{h.c.}) 
\end{split}  
+h\sum_{i=1}^N (2a_i^\dag a_i-1) 
\end{equation}
 as $H = \mathbf{a}^\dag\mathbb{H}\mathbf{a}$ where $\mathbf{a}$ is a $2N$ dimensional column-vector $\mathbf{a} = (a_1,...,a_N,a_1^\dag,...,a_N^\dag)^T$ and $\mathbf{a}^\dag = (a_1^\dag,...,a_N^\dag,a_1,...,a_N)$.
Then, we can write $\mathbb{H}$ in block matrix form,
\begin{align}
    \mathbb{H} = \begin{pmatrix}
        A & B\\
        -B^* & -A^*
    \end{pmatrix}
\end{align}
As we're dealing with an Ising model, we can take the $N \times N$ size matrices $A$ and $B$ real. Here, the imposed periodic boundary conditions on our initial spins may translate both as antiperiodic and periodic boundary conditions (ABC and PBC) to our free fermions according to fermion parity. As the ground state of our model is in the ABC sector, we continue with ABC~\cite{PhysRevB.53.8486}. However, we also note that numerical simulations both with PBC- and ABC-imposed Bogoliubov matrices gave nearly indistinguishable results, which can be attributed to the $0$-mean and low-variance Gaussian distribution of $J_i$

Then, by simple algebra, we get,
\begin{align}
    A_{i,i} = h, A_{i,i+1} = A_{i+1,i} = -J_{i}/2\\
    B_{i,i+1} = -B_{i+1,i} = -J_{i}/2
\end{align}
for $i\leq N$ and $i=N+1 \equiv 1$.

To obtain the eigenvectors and eigenvalues of $\mathbb{H}$, we must solve the equation,
\begin{align}
    \begin{pmatrix}
        A & B\\
        -B & -A
    \end{pmatrix}\begin{pmatrix}
        \mathbf{u_\mu} \\
        \mathbf{v_\mu}
    \end{pmatrix} = \epsilon_\mu \begin{pmatrix}
        \mathbf{u_\mu} \\
        \mathbf{v_\mu}
    \end{pmatrix}
\end{align}
When we organize the eigenvectors $\mathbf{u_\mu}$ and $\mathbf{v_\mu}$ into a block matrix $\mathbb{U}$ such that $\mathbb{U}^\dag\mathbb{H}\mathbb{U} = \text{diag}(\epsilon_\mu,-\epsilon_\mu)$, we get,
\begin{align}
    \mathbb{U} = \begin{pmatrix}
        \mathbf{u}_1 & ... & \mathbf{u_N} & \mathbf{v}_1 & ... & \mathbf{v}_N \\
        \mathbf{v}_1 &... &\mathbf{v}_N & \mathbf{u}_1 & ... & \mathbf{u}_N
    \end{pmatrix}
\end{align}
Now, with the aid of matrix $\mathbb{U}$, we can define new fermionic operators $b_i$ through a unitary transformation on $a_i$,
\begin{equation}\mathbf{b} = \mathbb{U}^\dag \mathbf{a}\end{equation}
Now, one can check that operators $b_i$ diagonalize our initial Hamiltonian $H$,
\begin{equation}\displaystyle H = \sum_{i=1}^N 2\epsilon_i(b_i^\dag b_i - \frac{1}{2})\end{equation}

Thus, we achieve diagonalization using the Bogoliubov transform.

\section{\label{app:exp}Calculating Hamiltonian Expectation Values}
To calculate the expectation value of our Hamiltonian at some state $A$, two pieces of information are required \cite{Quan_2007}:
\begin{itemize}
    \item First, we need the energy levels $E_j$ of the Hamiltonian. Through the Bogoliubov transform, these are equivalent to $\epsilon_j$.
    \item Second, we need the probability distribution that the system follows. As our model is a fermionic model, the system will follow a Fermi-Dirac distribution, modeled by the function:
    \begin{equation}f(E,T) = (e^{E/T}+1)^{-1}\end{equation}
    where $T$ is the temperature of the system while $E$ denotes the energy eigenvalue of a specific state.
\end{itemize}
Note that during adiabatic strokes, the probability distribution of the system is invariant while the energy levels change. On the contrary, during thermalization, the energy levels are invariant while the probability distribution changes. 
While finding the energy difference of the system before and after thermalization, then, we must use the same energy eigenvalues but different probability distributions.
For example, denote the energy levels of the Hamiltonian at $h(t) = h_i$ with $\epsilon_i$ while denoting the energy levels at $h(t) = h_f$ with $E_i$. Then, if we want to calculate the energy difference of the system before and after thermalization with the hot bath at $h(t) = h_f$, we get,
\begin{equation}\Delta\langle H \rangle = \sum_i E_i(f(E_i,T_h)-f(\epsilon_i,T_c))\end{equation}
In our study, we use the Bogoliubov eigenvalues to study the relevant thermodynamic quantities following~\cite{Piccitto_2022}, as the general behavior of the thermodynamic quantities. is similar to the real calculations using actual eigenenergies.

\bibliography{BibTexSpinGlass}

\end{document}